\documentclass[a4,11pt]{article}
\usepackage{graphicx} 
\usepackage{float}

\usepackage{amsmath,amsthm}
\usepackage{amsfonts}
\usepackage{latexsym}
\usepackage{amssymb}
\usepackage{setspace}
\usepackage{comment}

\makeatletter
 
  \@addtoreset{equation}{section}
 \makeatother

\begin{document}
\title{Asymptotic Expansion as Prior Knowledge in\\
Deep Learning Method for high dimensional BSDEs}


\author{
Masaaki Fujii\footnote{Quantitative Finance Course, Graduate School of Economics, The University of Tokyo.},
~~~~Akihiko Takahashi\footnote{Quantitative Finance Course, Graduate School of Economics, The University of Tokyo. },  
~~~~Masayuki Takahashi\footnote{Quantitative Finance Course, Graduate School of Economics, The University of Tokyo.}} 

\date{ First version: October 19, 2017\\
This version: March 5,  2019}
\maketitle



\newtheorem{definition}{Definition}[section]
\newtheorem{assumption}{Assumption}[section]
\newtheorem{condition}{$[$ C}
\newtheorem{lemma}{Lemma}[section]
\newtheorem{proposition}{Proposition}[section]
\newtheorem{theorem}{Theorem}[section]
\newtheorem{remark}{Remark}[section]
\newtheorem{example}{Example}[section]
\newtheorem{corollary}{Corollary}[section]
\def\n{{\bf n}}
\def\A{{\bf A}}
\def\B{{\bf B}}
\def\C{{\bf C}}
\def\D{{\bf D}}
\def\E{{\bf E}}
\def\F{{\bf F}}
\def\G{{\bf G}}
\def\H{{\bf H}}
\def\I{{\bf I}}
\def\J{{\bf J}}
\def\K{{\bf K}}
\def\L{{\bf L}}
\def\M{{\bf M}}
\def\N{{\bf N}}
\def\O{{\bf O}}
\def\P{{\bf P}}
\def\Q{{\bf Q}}
\def\R{{\bf R}}
\def\S{{\bf S}}
\def\T{{\bf T}}
\def\U{{\bf U}}
\def\V{{\bf V}}
\def\W{{\bf W}}
\def\X{{\bf X}}
\def\Y{{\bf Y}}
\def\Z{{\bf Z}}
\def\cala{{\cal A}}
\def\calb{{\cal B}}
\def\calc{{\cal C}}
\def\cald{{\cal D}}
\def\cale{{\cal E}}
\def\calf{{\cal F}}
\def\calg{{\cal G}}
\def\calh{{\cal H}}
\def\cali{{\cal I}}
\def\calj{{\cal J}}
\def\calk{{\cal K}}
\def\call{{\cal L}}
\def\calm{{\cal M}}
\def\caln{{\cal N}}
\def\calo{{\cal O}}
\def\calp{{\cal P}}
\def\calq{{\cal Q}}
\def\calr{{\cal R}}
\def\cals{{\cal S}}
\def\calt{{\cal T}}
\def\calu{{\cal U}}
\def\calv{{\cal V}}
\def\calw{{\cal W}}
\def\calx{{\cal X}}
\def\caly{{\cal Y}}
\def\calz{{\cal Z}}
%
\def\sskip{\hspace{0.5cm}}
\def\simleq{ \raisebox{-.7ex}{\em $\stackrel{{\textstyle <}}{\sim}$} }
\def\leqsim{ \raisebox{-.7ex}{\em $\stackrel{{\textstyle <}}{\sim}$} }
\def\ep{\epsilon}
\def\half{\frac{1}{2}}
\def\iku{\rightarrow}
\def\Iku{\Rightarrow}
\def\ikup{\rightarrow^{p}}
\def\inclusion{\hookrightarrow}
\def\cadlag{c\`adl\`ag\ }
\def\up{\uparrow}
\def\down{\downarrow}
\def\doti{\Leftrightarrow}
\def\douti{\Leftrightarrow}
\def\dochi{\Leftrightarrow}
\def\douchi{\Leftrightarrow}%
\def\yy{\\ && \nonumber \\}
\def\y{\vspace*{3mm}\\}
\def\nn{\nonumber}
\def\be{\begin{equation}}
\def\ee{\end{equation}}
\def\bea{\begin{eqnarray}}
\def\eea{\end{eqnarray}}
\def\beas{\begin{eqnarray*}}
\def\eeas{\end{eqnarray*}}
%
\def\hd{\hat{D}}
\def\hv{\hat{V}}
\def\hsd{{\hat{d}}}
\def\hx{\hat{X}}
\def\hsx{\hat{x}}
\def\bsx{\bar{x}}
\def\bsd{{\bar{d}}}
\def\bx{\bar{X}}
\def\ba{\bar{A}}
\def\bb{\bar{B}}
\def\bc{\bar{C}}
\def\bv{\bar{V}}
\def\balpha{\bar{\alpha}}
\def\bbalpha{\bar{\bar{\alpha}}}
\def\combi{\l(\begin{array}{c}\alpha\\ \beta \end{array}\r)}
\def\f{^{(1)}}
\def\s{^{(2)}}
\def\ss{^{(2)*}}
\def\l{\left}
\def\r{\right}
\def\a{\alpha}
\def\b{\beta}
\def\L{\Lambda}


\def\E{{\bf E}}

\def\cadlag{{c\`adl\`ag~}}

\def\calf{{\cal F}}
\def\wt{\widetilde}
\def\mbb{\mathbb}
\def\ol{\overline}
\def\ul{\underline}
\def\sign{{\rm{sign}}}
\def\wh{\widehat}
\def\tp{{t^\prime}}
\def\sp{{s^\prime}}
\def\xp{{x^\prime}}
\def\bq{{\bar{q}}}
\def\p{\prime}
\def\ep{\epsilon}
\def\del{\delta}
\def\Del{\Delta}
\def\part{\partial}

\def\nn{\nonumber}
\def\be{\begin{equation}}
\def\ee{\end{equation}}
\def\bea{\begin{eqnarray}}
\def\eea{\end{eqnarray}}
\def\beas{\begin{eqnarray*}}
\def\eeas{\end{eqnarray*}}
\def\l{\left}
\def\r{\right}

\def\bull{$\bullet~$}

\newcommand{\Slash}[1]{{\ooalign{\hfil/\hfil\crcr$#1$}}}
\vspace{0mm}

\begin{abstract}
We demonstrate that the use of asymptotic expansion as prior knowledge in the {\it ``deep BSDE solver"},
which is a deep learning method for high dimensional BSDEs proposed by Weinan E, Han \& Jentzen (2017),
drastically reduces the loss function and accelerates the speed of convergence.
We illustrate the technique and its implications by using Bergman's model with different lending and borrowing rates
as a typical model for FVA as well as a class of solvable BSDEs with quadratic growth drivers.
We also present an extension of the deep BSDE solver for reflected BSDEs representing American option prices.
\end{abstract}

\section{Introduction}
This work presents a simple acceleration technique for deep learning methods
for high-dimensional backward stochastic differential equations (BSDEs).
Since its invention, BSDEs have attracted many mathematicians by 
their deep connections to non-linear partial differential equations and stochastic control problems.
The relevance of BSDEs for financial problems has also increased recently, particularly since
the financial crisis. Early attempts such as Fujii \& Takahashi (2012, 2013)~\cite{FT-CDS, FT-CVA} 
and Crepey (2015)~\cite{Crepey-TVA} have shown that 
BSDEs are  indispensable tools to describe the non-linear effects in various valuation adjustments 
stemming from collateralization,  credit risks, funding and regulatory costs.  
See Brigo, Morini \& Pallavicini (2013)~\cite{Brigo-book} and Crepey \& Bielecki (2014)~\cite{Crepey-book} as  reviews
on the financial problems closely related to BSDEs.
Due to their direct connection to the 
stochastic control problems, BSDEs are also actively studied to obtain probabilistic representation of 
optimal liquidation, switching and portfolio selection  problems.

The progress in numerical computation schemes for BSDEs has also been significant.
The famous $\mbb{L}^2$-regularity established by Zhang (2001, 2004)~\cite{JZhangPHD,JZhang}
was soon followed by now standard regression-based Monte-Carlo simulation scheme developed, among others, by
Bouchard \& Touzi (2004)~\cite{Bouchard-Touzi}, Gobet, Lemor \& Warin (2005)~\cite{Gobet-Lemor-Warin}.
There now exist many extensions of these fundamental works  to various types of BSDEs.
Unfortunately, however, applications to high dimensional problems required in the practical setups have been almost infeasible
due to their heavy numerical burden. 
This is the biggest obstacle that has been hindering BSDE from becoming a standard mathematical tool in the financial
industry. As a result, many of the XVAs are usually implemented as linear approximations by neglecting
sometimes crucial non-linear feedback effects.

A potential breakthrough may come from the recent boom as well as explosive progress of reinforcement machine learning,
which makes use of deep neutral networks mimicking the cognitive mechanism of human brains. 
In fact, Weinan E, Han \& Jentzen (2017)~\cite{Jentzen}
motivated by the work of Weinan E \& Han (2016)~\cite{E-Han} have just demonstrated 
astonishing power of ``deep BSDE solver" for high dimensional problems,
which is based on the deep neural networks constructed by the free package {\it Tensorflow}.\footnote{See also Beck, E \& Jentzen (2017)~\cite{Jentzen-2ndOrder},  where the method is applied to  different types of BSDEs, 
and \cite{Jentzen-Kruse-1, Jentzen-Kruse-2} as different approaches to high-dimensional problems.}
Their main idea is to interpret a Markovian BSDE 
\be
Y_t=\Phi(X_T)+\int_t^T f(s,X_s,Y_s,Z_s)ds-\int_t^T Z_s dW_s,~t\in[0,T] \nn
\ee
as a control problem minimizing the square difference $|\Phi(X_T)-\wh{Y}_T|^2$. 
Here,  $\Phi(X_T)$ is the terminal condition and $\wh{Y}_T$ the terminal 
value of forwardly simulated process $(Y_t)_{t\in[0,T]}$ based on
the estimated initial value $Y_0$  as well as the coefficients of Brownian motions $(Z_t)_{t\in[0,T]}$,
which are treated as the {\it control} variables in the minimization problem.
This is nothing but concrete implementation of {\it Method of Optimal Control} as a solution technique for FBSDEs
given in Chapter 3 of Ma \& Yong (2000)~\cite{Ma-Yong}. 
Although the mathematical understanding for the deep learning algorithm is still in its infancy, the deep BSDE solver seems to be capable of handling  high dimensional problems very efficiently in a quite straightforward way.\footnote{As interesting applications of machine learning to various investment strategies, 
see Nakano et.al. (2017)~\cite{Nakano-Fuzzy,
Nakano-Fuzzy-2, Nakano-3}.}

Despite its remarkable success for high dimensional problems, it is not free from some important issues.
By closely studying the deep BSDE solver given in \cite{Jentzen}, we find that 
its direct application to the Bergman (1995)~\cite{Bergman} equation, as a typical model for funding value adjustment (FVA)
with different lending and borrowing rates, yields persistently high loss 
function $|\Phi(X_T)-\widehat{Y}_T|^2$
even when the estimated $Y_0$, which corresponds to the price of the contingent claim, 
is quite accurate. The slow convergence and large loss function seem to arise from 
non-smooth terminal condition $\Phi(\cdot)$ as well as driver $f(\cdot)$ of the BSDE, which are ubiquitous in financial applications.

Let us denote by $Y_t^{t,x}$ the value of the solution at time $t$ under the $t$-initialized setup 
with the input $X_t=x$. Due to the strong Markov property of Brownian motions and 
Blumenthal's $0$-$1$ law, $Y_t^{t,x}$ is almost surely deterministic and hence is given by some Borel-measurable 
function $u$ as $Y_t^{t,x}=u(t,x)$. When $u$ is smooth, it is well-known that
$\bigl(Z_s^{t,x}=\part_x u(s,X_s)\sigma(s,X_s),~s\geq t\bigr)$ where $\sigma(s,X_s)$ is the diffusion coefficient of the SDE 
specifying the dynamics of the forward component $(X_s)_{s\geq t}$.
This observation shows  that the deep BSDE solver 
is actually looking for the optimal delta-hedging strategy $\part_x u(t,x)$ under given time partition. It is thus natural to have slower convergence when the relevant coefficient functions
are not smooth. Similarly, from the financial viewpoint, the loss function
is the square of ``replication error" from the associated delta-hedging strategy.
Therefore, even when $Y_0$ is known to be accurate, the resultant strategy is  not useful when the loss function
remains high. Worse,  we do not know the accurate value of $Y_0$ in general.
Only available criterion at hand is the famous stability result of BSDEs to guarantee the uniqueness of their solutions,
which thus requires the convergence of the loss function to a sufficiently small value.
Since many of the financial problems related to the valuation adjustments i.e. XVAs 
have quite similar form to the Bergman's model, this is not an exceptional problem.
The speed of learning process is found to be strongly affected  by the correlation among the underlying security processes, too.

It has been widely known that the prior knowledge to prepare the starting point of the learning process significantly 
affects the performance of deep learning methods. In this work, we demonstrate that a simple approximation formula
based on an asymptotic expansion (AE)
of BSDEs serves as  very efficient prior knowledge for the deep BSDE solver.
Using the method proposed in the works Fujii \& Takahashi (2012, 2019)~\cite{FT-analytic, FT-AE}, one obtains an analytic expression of 
approximate $Z^{\rm AE}$. We write $Z=Z^{\rm AE}+Z^{\rm Res}$ and apply the reinforcement learning only to the residual term 
$Z^{\rm Res}$. We shall show that the use of $Z^{\rm AE}$ drastically reduces the loss function and accelerates 
the speed of convergence.
We have also presented an extension of the deep BSDE solver for reflected backward stochastic differential equations 
(RBSDEs), which become relevant when studying optimal stopping/switching problems.
Using an American basket option as an example, we have shown that the effectiveness of AE
as prior knowledge still holds for RBSDEs.
Numerical examples for a class of quadratic growth BSDEs (qg-BSDEs) are also given.
Despite the notorious difficulty to obtain stable numerical results for qg-BSDEs,
the deep BSDE solver with AE is shown to handle the problem quite efficiently.

\section{An application to Bergman's model}
\subsection{Model}
\label{sec-Bergman-model}
Let us consider the filtered probability space $(\Omega,\calf_T, \mbb{F}=(\calf_t)_{t\in[0,T]},\mbb{P})$ 
generated by a $d$-dimensional Brownian motion $(W^\alpha)_{\alpha=1}^d$, which is assumed to satisfy the usual conditions.
We suppose that the $d$ risky assets follow the dynamics
\bea
X_t^i=x_0^i+\int_0^t \mu^i X_s^i ds+\int_0^t X_s^i \sigma^i \sum_{\alpha=1}^d \rho_{i,\alpha}dW_s^\alpha, ~t\in[0,T], i=1,\cdots, d 
\label{eq-X}
\eea
where $x_0^i>0$ is the initial value, $\mu^i, \sigma^i>0$ are constants
and $(\rho_{i,\alpha})_{i,\alpha=1}^d$ is the square root of the (instantaneous) correlation matrix among $X^i$,
normalized as $(\rho\rho^\top)_{i,i}=1,~i=1,\cdots,d$. $\rho$ is assumed to be invertible.
There are two interest rates, one is for lending $r>0$ and  the other $R>r$  for borrowing.
The dynamics of portfolio value $(Y_t)_{t\in[0,T]}$ under the least-borrowing self-financing strategy for replicating 
the terminal payoff $\Phi(X_T)$, where $\Phi:\mbb{R}^d\rightarrow \mbb{R}$ is a Lipschitz continuous function, is given by 
the following BSDE:
\bea
Y_t&=&\Phi(X_T)-\int_t^T \Bigl\{ rY_s+\sum_{i,\alpha=1}^d Z_s^\alpha(\rho^{-1})_{\alpha,i}\frac{\mu^i-r}{\sigma^i}
-\Bigl(\sum_{i,\alpha=1}^d Z_s^\alpha(\rho^{-1})_{\alpha,i}\frac{1}{\sigma^i}-Y_s\Bigr)^+(R-r)\Bigr\}ds\nn \\
&&-\int_t^T \sum_{\alpha=1}^d Z_s^\alpha dW_s^\alpha, \quad t\in[0,T]~.
\label{eq-Bergman-BSDE}
\eea
See, for example, Example 1.1 in \cite{ElKaroui-Review} as a simple derivation of the above form.
The existence of unique solution is guaranteed by the standard results for the Lipschitz BSDEs.
Note that the cash amount invested to the ith risky asset at time $t$
 is  given by $\pi_t^i=\sum_{\alpha=1}^d Z_t^\alpha(\rho^{-1})_{\alpha,i}/\sigma^i$.

\begin{remark}
When one applies the method to a BSDE, 
the coefficient of the Brownian motion $Z$ is estimated for each scenario at each time step.
By defining the deterministic map $u:[0,T]\times \mbb{R}^d\rightarrow \mbb{R}$ by
$u(t,x):=Y_t^{t,x}$, where $(t,x)$ denotes the initial data of the underlying security price process $X$,
the representation theorem of the BSDE implies (when $u$ has appropriate regularity) that
$Z_s^{\alpha}=\sum_{i=1}^d \part_{x^i}u(s,X_s)\sigma^i X_s^i \rho_{i,\alpha}$.
In other words, one obtains not only the price but also the path-wise delta sensitives through the learning process
of the deep BSDE solver.
This is quite valuable information, for example, to estimate the necessary independent amount based on the standard initial margin 
method (SIMM).
\end{remark}

\subsection{Asymptotic expansion based on driver's linearization}
We adopt an asymptotic expansion method proposed in \cite{FT-analytic} which is based on 
perturbative expansion of the non-linear driver of the BSDE around a linear term. Mathematical
justification of the expansion is available in Takahashi \& Yamada (2015)~\cite{Takahashi-Yamada}.
Its implementation with a particle method Fujii \& Takahashi (2015)~\cite{FT-particle} has been successfully 
applied to large scale numerical simulations in many works using up to the second order expansions.
See, for example,  Crepey \& Nguyen (2016)~\cite{Crepey-Nguyen} and references therein.

In this work, we only use the leading term of the asymptotic expansion. For higher order corrections, see discussions
and examples available in \cite{FT-analytic, Takahashi-Yamada}.  
According to \cite{FT-analytic}, we consider (\ref{eq-Bergman-BSDE})
as the perturbed model around the linear driver:
\bea
Y_t^{\ep}&=&\Phi(X_T)-\int_t^T \Bigl\{ rY_s^\ep+\sum_{i,\alpha=1}^d Z_s^{\alpha,\ep}(\rho^{-1})_{\alpha,i}\frac{\mu^i-r}{\sigma^i}
-\ep \Bigl(\sum_{i,\alpha=1}^d Z_s^{\alpha,\ep}(\rho^{-1})_{\alpha,i}\frac{1}{\sigma^i}-Y_s^\ep\Bigr)^+(R-r)\Bigr\}ds\nn \\
&&-\int_t^T \sum_{\alpha=1}^d Z_s^{\alpha,\ep} dW_s^\alpha, \quad t\in[0,T]~. \nn
\eea
The idea of the approximation is to expand $(Y^{\ep},Z^{\alpha,\ep})$ around $\ep=0$.
The leading order terms $(Y^{(0)},Z^{\alpha,(0)}):=(Y^{\ep},Z^{\alpha,\ep})|_{\ep=0}$ are determined by
\bea
Y_t^{(0)}&=&\Phi(X_T)-\int_t^T \Bigl\{ rY_s^{(0)}+\sum_{i,\alpha=1}^d Z_s^{\alpha,(0)}(\rho^{-1})_{\alpha,i}\frac{\mu^i-r}{\sigma^i}
\Bigr\}ds-\int_t^T \sum_{\alpha=1}^d Z_s^{\alpha,(0)} dW_s^\alpha, \quad t\in[0,T]~. \nn
\eea
This immediately gives $Y_t^{(0)}=e^{-r(T-t)}\mbb{E}^{\mbb{Q}}\Bigl[\Phi(X_T)|\calf_t\Bigr], ~t\in[0,T]$ 
with the probability measure $\mbb{Q}$ defined by
\bea
\frac{d\mbb{Q}}{d\mbb{P}}=\cale\Bigl(-\int_0^T \sum_{\alpha, j=1}^d(\rho^{-1})_{\alpha,j}\frac{\mu^j-r}{\sigma^j} dW_s^\alpha\Bigr)\nn 
\eea
where $\cale$ is Dol\'eans-Dade exponential. Since $Y^{(0)}$ is equal to the price process in Black-Scholes model 
with the risk-free rate $r$,
$Z^{(0)} (=:Z^{\rm AE})$ is obtained as deltas with respect to $X$ multiplied by the diffusion coefficient $\sigma^i X^i$.
For example, if $d=1$ and $\Phi(X_T)=\max(X_T-K,0)$, one has $Z^{(0)}_t=N(d_+)\sigma X_t$ 
where $N(\cdot)$ is the distribution function of the standard normal and 
$d_+=\frac{1}{\sigma\sqrt{T-t}}\Bigl(\log\Bigl(\frac{F_t}{K}\Bigr)+\frac{1}{2}\sigma^2 (T-t)\Bigr), \quad F_t=e^{r(T-t)}X_t$.

\begin{remark}
We should emphasize that an analytical expression can be obtained even when $\Phi$ and the process $X$
have more general forms. This is a well-known application of asymptotic expansion technique to European contingent claims.
See Takahashi (2015)~\cite{Takahashi-review} and references therein for details on this topic.
It is also important to notice that the leading order asymptotic expansion keeps linearity.
Since it is derived from a linearized BSDE, the resultant approximation is also linear with respect to the cash flow.
Thus, in particular,  the approximation for a derivatives portfolio is given by a sum of $Z^{\rm AE}$ for its
individual contract.
\end{remark}
In the following, in order to focus on the implications of AE as prior knowledge in the deep BSDE solver
instead of deriving AE formulas for general setups, we only deal with the terminal conditions consisting of call/put options
and the log-normal process for $X$ in (\ref{eq-X}).

\subsection{Numerical examples}
\subsubsection{Purely call terminals}
Suppose that the terminal condition is given by
\be
\Phi(X_T)=\sum_{i=1}^d q^i \max(X_T^i-K^i,0), \quad q^i>0, ~i=1,\cdots,d. \nn
\ee
In this case, the one who tries to replicate the terminal payoff must always hold a long position for every risky asset.
Since this implies that she must always borrow cash to finance her  hedging position, the BSDE (\ref{eq-Bergman-BSDE}) is equivalent to
\bea
Y_t&=&\Phi(X_T)-\int_t^T \Bigl\{ RY_s+\sum_{i,\alpha=1}^d Z_s^\alpha(\rho^{-1})_{\alpha,i}\frac{\mu^i-R}{\sigma^i}\Bigr\}ds-\int_t^T \sum_{\alpha=1}^d Z_s^\alpha dW_s^\alpha \nn 
\eea
Notice that this holds true irrespective of the correlation among $X$'s. After a simple measure change, one sees that 
the exact solution of $Y_0$ is given by the corresponding Black-Scholes formula with $r$ replaced by $R$.

\begin{figure}[h]
\hspace{-9mm}
\includegraphics[width=85mm]{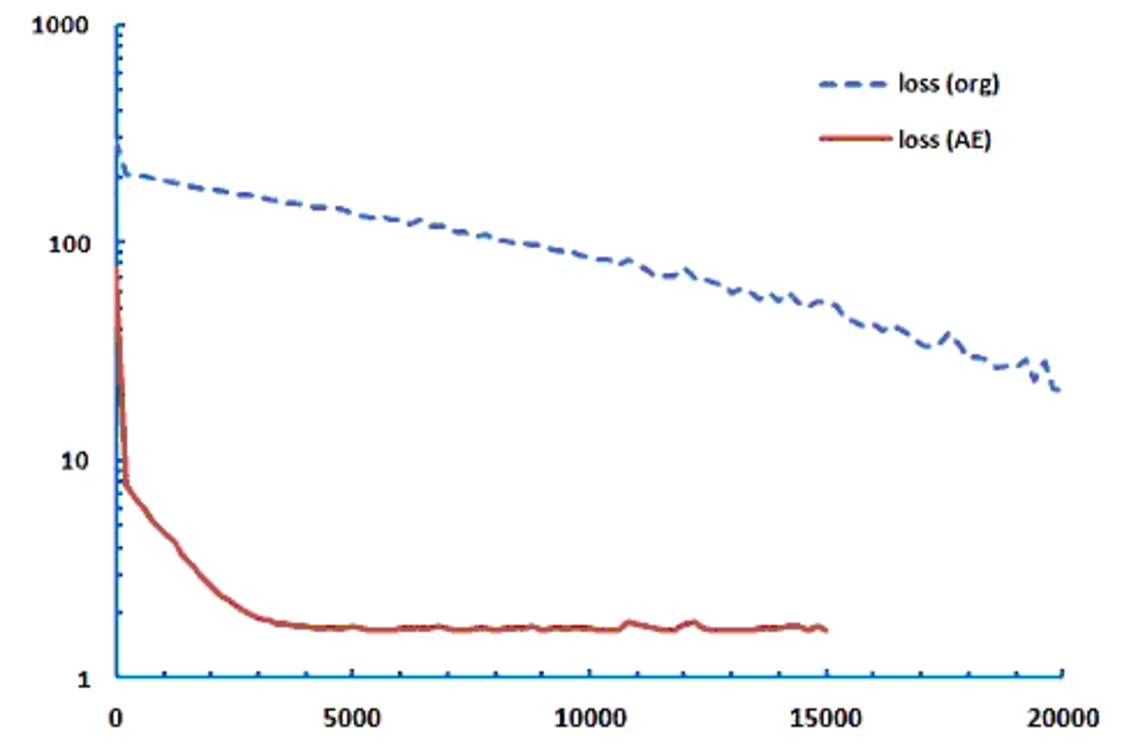}
\includegraphics[width=85mm]{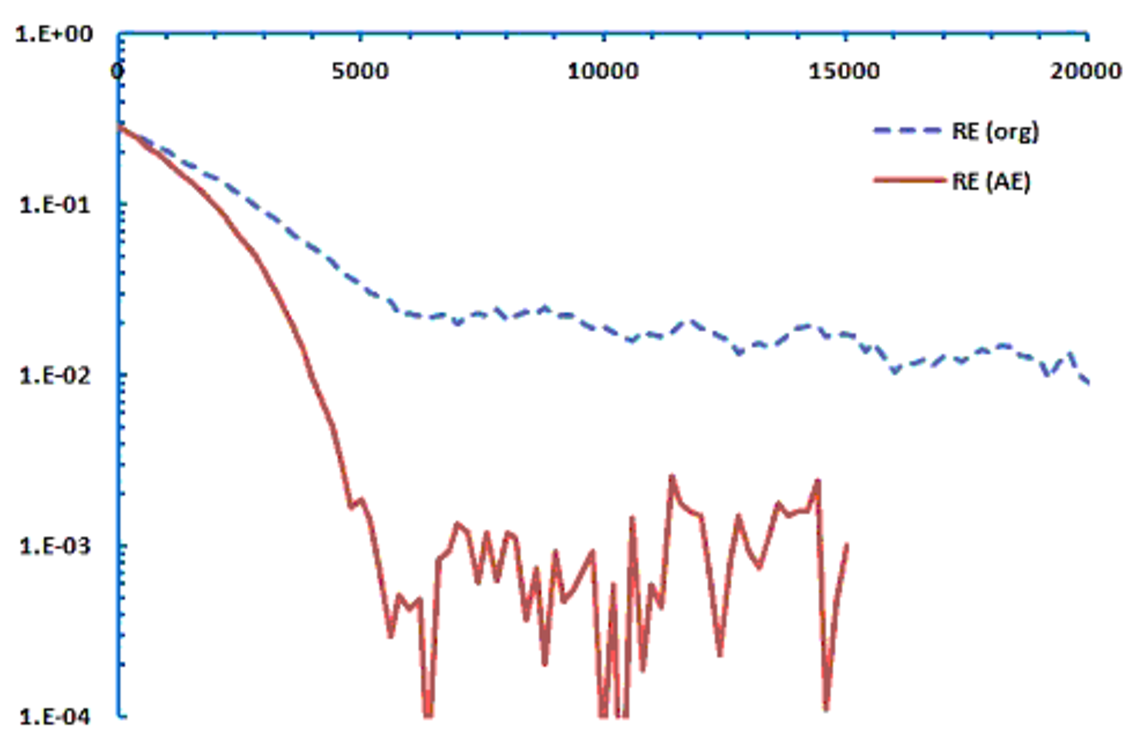}
\caption{\small Comparison of the loss function and the relative error w.r.t. the exact solution between
the direct use of the deep BSDE solver (indicated by a dashed line) and the one using AE
as prior knowledge (indicated by a solid line) for the 1-dimensional case. The horizontal axis is the number of iteration steps of the learning process. }
\label{Fig-d1Call}
\end{figure}

Let us start from the simplest one-dimensional example with
${\rm set~A}:=\{d=1,q=1,\mu=0.05, r=0.01,   R=0.06, \sigma=0.3, T=0.5, K=103\}$. In this case, the above discussion gives $Y_0=8.4672$ as
the exact solution. We have used ${\rm n\_{time}}=50$ (time discretization), ${\rm batch\_size}=64$, ${\rm n\_layer}=4$, 
${\rm learning\_rate}=10^{-3}$ in the deep BSDE solver~\cite{Jentzen} in Figure~\ref{Fig-d1Call}.
The loss function is estimated with 1024 paths.
As explained before,  we have used only the leading order approximation as $Z^{\rm AE}$ and put $Z=Z^{\rm AE}+Z^{\rm Res}$
in the deep BSDE solver, where only the residual term $Z^{\rm Res}$ and $Y_0$ are used as the targets of the training process.
In Figure~\ref{Fig-d1Call}, we have compared the performance of the deep BSDE solver with and without AE as prior knowledge.
It is observed that one achieves much quicker convergence and roughly by one order of magnitude smaller relative error
when one uses AE as prior knowledge. After roughly 5,000 iterations,
its loss function reaches 1.7.  Since the option is around at-the-money, the gamma at the last stage
is huge in many paths. If the delta-hedging at the last period $\Del t=1/100$ completely fails, its contribution to the 
loss function is estimated roughly by $(100\times 0.3\times \sqrt{\Del t}\times 0.5)^2=2.25$. 
This estimate implies that the deep BSDE solver with AE reaches its limit performance already at 5,000 iterations.
When AE is not used, one sees that the loss function (and hence the replication error) remains rather big and slow to converge.

Notice that the deep BSDE solver uses {\bf{tf.train.AdamOptimizer}} available in the TensorFlow package 
for optimizing the coefficient matrices usually denoted by {\bf{w}}.
This is the algorithm proposed in \cite{Adam}, in which the learning\_rate $10^{-3}$ is recommended as a default value.
Although one can speed up the learning process by increasing the learning rate, this is not always appropriate.
Let us study the effects of the learning rate using the next 30-dimensional example.

\begin{figure}[h]
\hspace{-9mm}
\includegraphics[width=85mm]{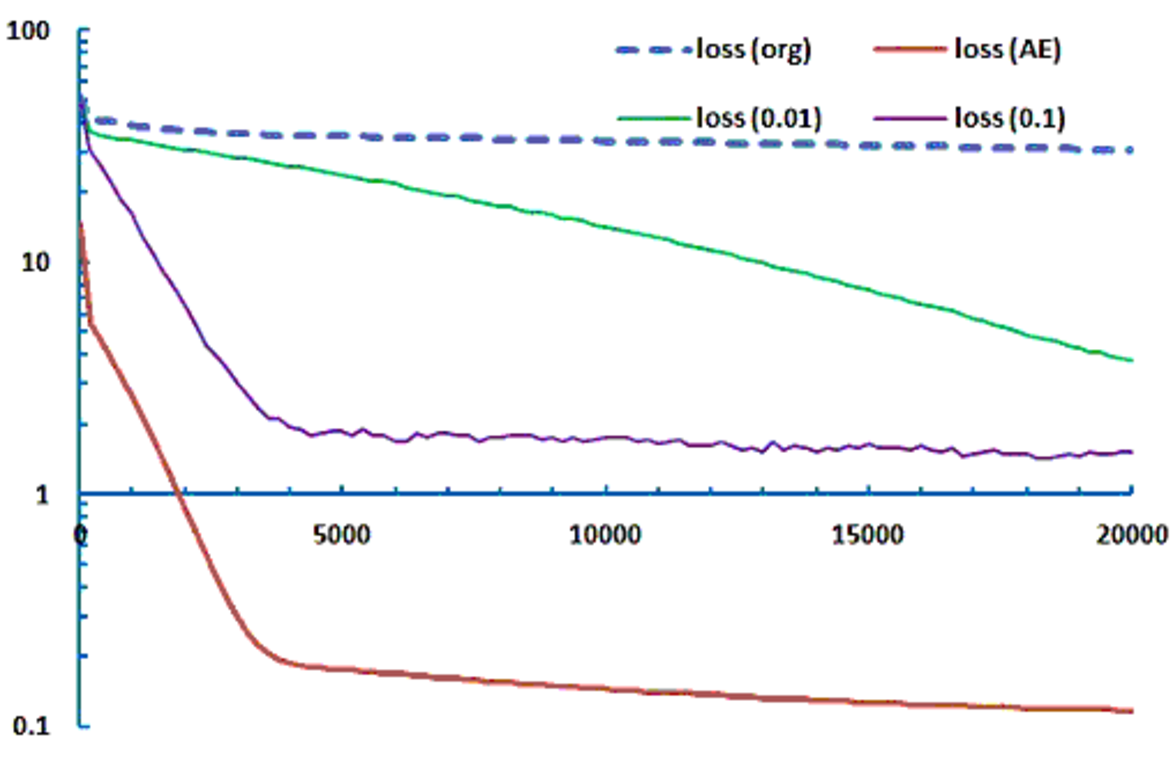}
\includegraphics[width=85mm]{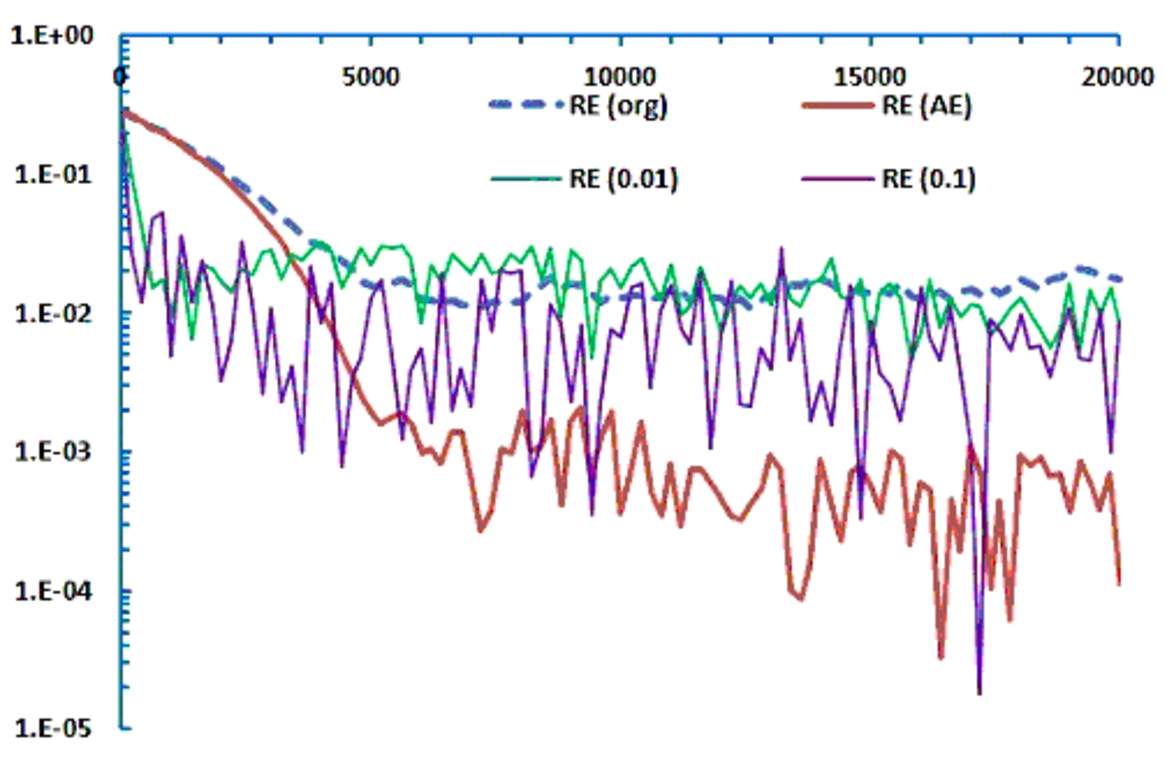}
\caption{\small Comparison of the loss function and the relative error w.r.t. the exact solution
with and without AE as prior knowledge for the 30-dimensional case with correlation implied by $\gamma=0.06$
in (\ref{eq-rho}). The two {\it thin} solid lines (green and purple) denote the cases without using AE but the learning\_rate
replaced by $0.01$ and $0.1$, respectively.
The horizontal axis is the number of iteration steps of the learning process. }
\label{Fig-d30Call}
\end{figure}

We study the case with $d=30$ where the set~A is replaced by $q^i=1/d$ and $\mu^i,\sigma^i, K^i,~i=1,\cdots d$
are common and the same with those in set~A i.e., $\Phi$ is given by the average of the call options.
The matrix $\rho$ is assumed to have the form
\be \rho=\frac{1}{\sqrt{1+(d-1)\gamma^2}}
\begin{pmatrix}
1 &~\cdots~& \gamma\\
\vdots   &~\ddots   &\vdots \\
\gamma & \cdots  &1
\end{pmatrix}
\label{eq-rho}
\ee
with $\gamma=0.06$. This implies that the correlation for every pair $(X^i,X^j)$ is about 20\%.
The exact value of $Y_0$ must be the same $Y_0=8.4672$.
In Figure~\ref{Fig-d30Call}, we have provided the numerical results for this case. 
In addition to those with the default learning\_rate $=10^{-3}$, 
we have added two cases with learning\_rate $=10^{-2}$ and $10^{-1}$ without AE as prior knowledge.
One sees, for example, learning\_rate $=10^{-1}$ yields  a fast decline of the loss function in the
first 5,000 steps comparable to the case with AE, but it stops at the level 10 times higher than the case with AE.
Moreover, the estimated $Y_0$ (and hence the relative error) exhibits strong instability.
The use of asymptotic expansion with the default learning rate yields a more stable and accurate estimate.
Notice that the instability associated with a higher learning rate is more prominent 
for lower dimensional problems. The 30-dimensional example we have just considered, the instability is somewhat mitigated 
by the diversification effects from the imperfectly correlated 30 assets.

\begin{remark}
Dynamically choosing the optimal learning rate is an important issue and, in fact, is a popular
topic for researchers on computation algorithms. 
At the moment, however, there exists no established rule and it looks to depend
on a specific problem under consideration. As we have seen above, the optimal choice  
depends on the dimension of the forward process $X$ as well as their correlation 
even if the form of the BSDE is the same.
In the reminder of the paper,  
we shall fix the learning rate to the default value $10^{-3}$ for the {\bf{tf.train.AdamOptimizer}}
unless explicitly stated otherwise.
\end{remark}

\begin{remark}
We have no intention to deny the possibility that the convergence speed  
can be improved by implementing some hyper-parameter optimization, for example, tuning the learning rate, batch size, 
and the number of layers etc.  However, this is not usually an easy task requiring trial and error.
Note that one may even use our acceleration technique together with the hyper-parameter optimization. 
\end{remark}

\subsubsection{Call spread}
\begin{figure}[h]
\hspace{-9mm}
\includegraphics[width=85mm]{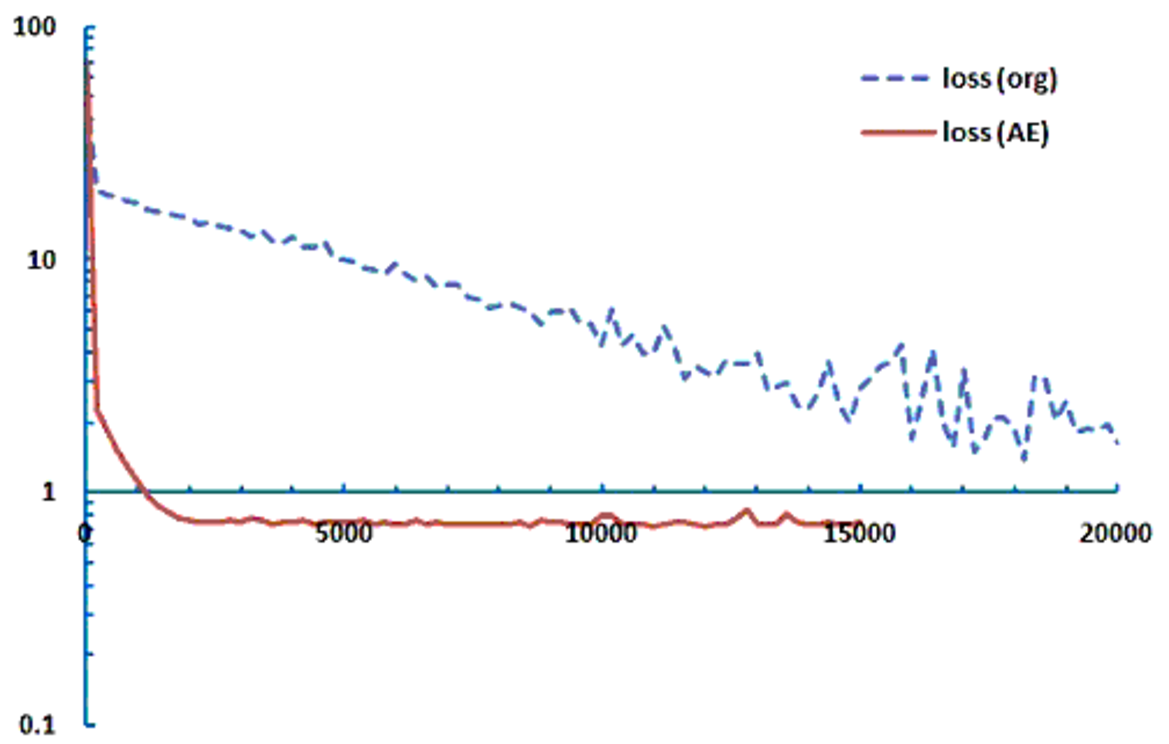}
\includegraphics[width=85mm]{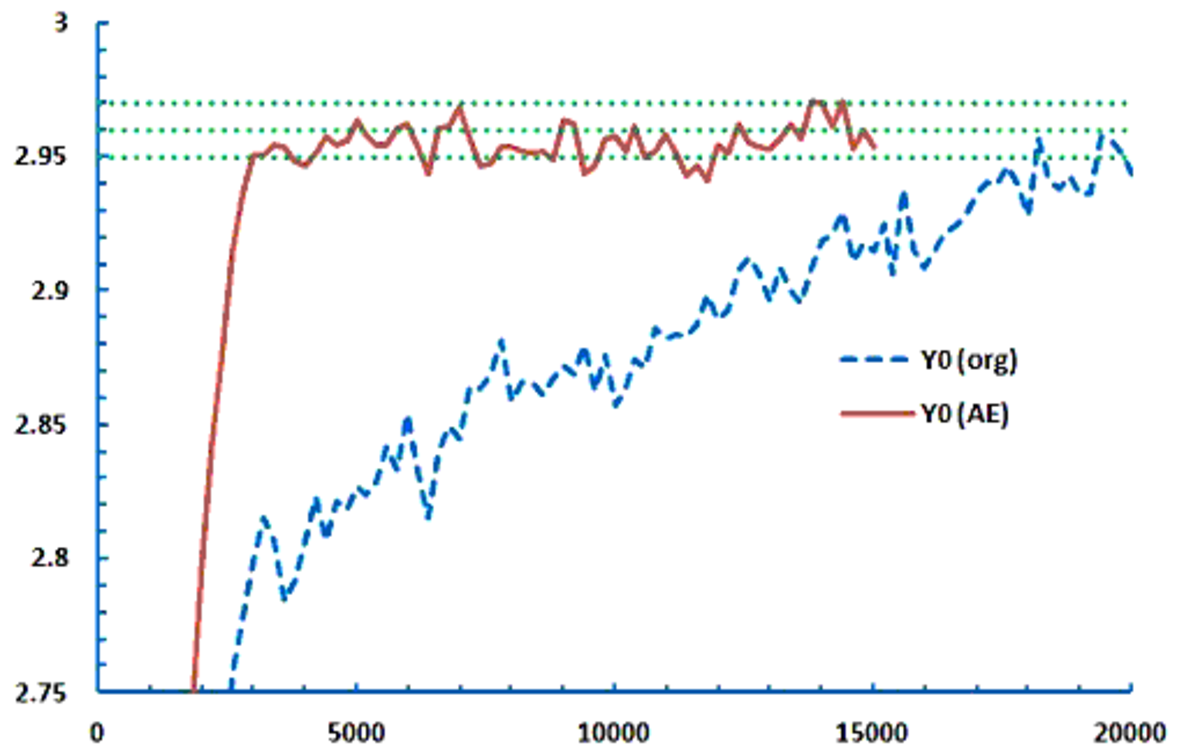}
\caption{\small Comparison of the loss function and the estimated $Y_0$ between
the direct use of the deep BSDE solver (indicated by a dashed line) and the one using AE
as prior knowledge (indicated by a solid line) for the 1-dimensional call spread. The three dotted lines denote $2.96\pm 0.01$
the value obtained in \cite{Bender-Steiner} using the regression-based Monte Carlo simulation. }
\label{Fig-d1CSP}
\end{figure}

We next  study the terminal function
$\Phi(X_T)=(X_T-K_1)^+-2(X_T-K_2)^+ \nn$
with $\{d=1, \mu=0.05, r=0.01, R=0.06, \sigma=0.2, T=0.25, K_1=95, K_2=105\}$.
For this 1-dimensional example of call spread, there is no closed-form solution anymore. 
However, it is estimated as $Y_0=2.96\pm 0.01$ in 
Bender \& Steiner (2012)~\cite{Bender-Steiner} using the regression based Monte Carlo scheme 
improved by the martingale basis functions.
The numerical results are given in Figure~\ref{Fig-d1CSP}. 
A much quicker convergence and smaller loss function are observed as before.

\begin{figure}[h]
\vspace{0mm}
\hspace{-9mm}
\includegraphics[width=85mm]{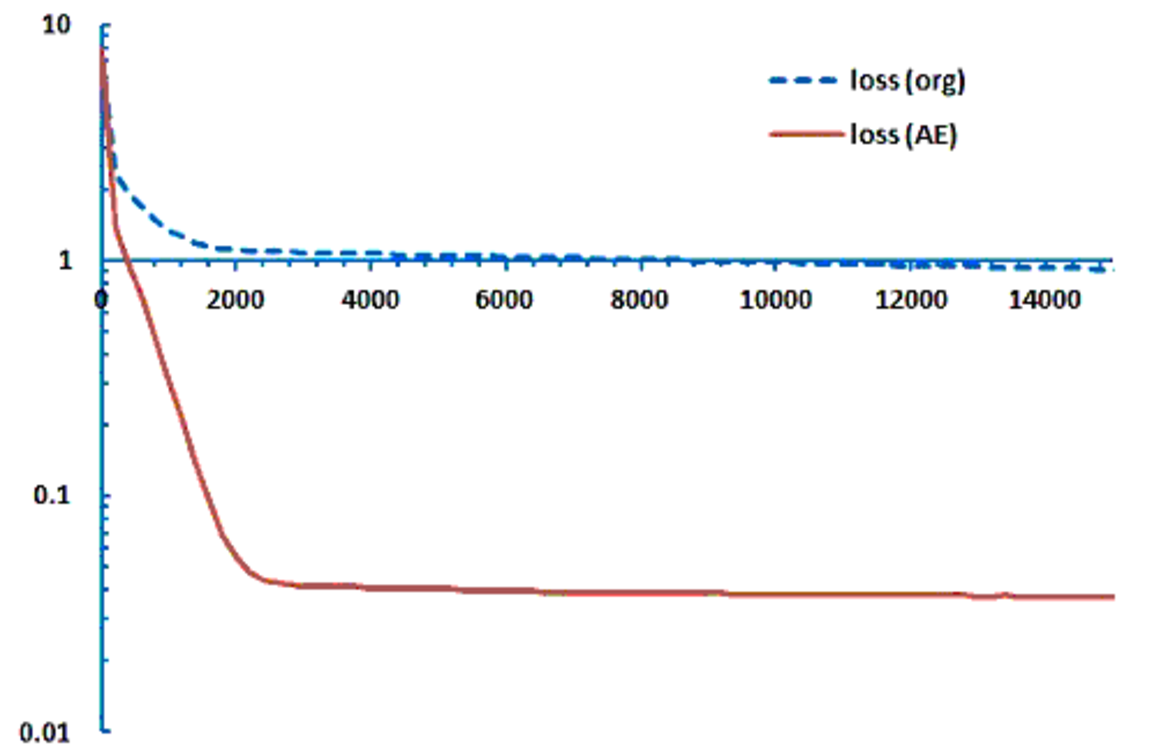}
\includegraphics[width=85mm]{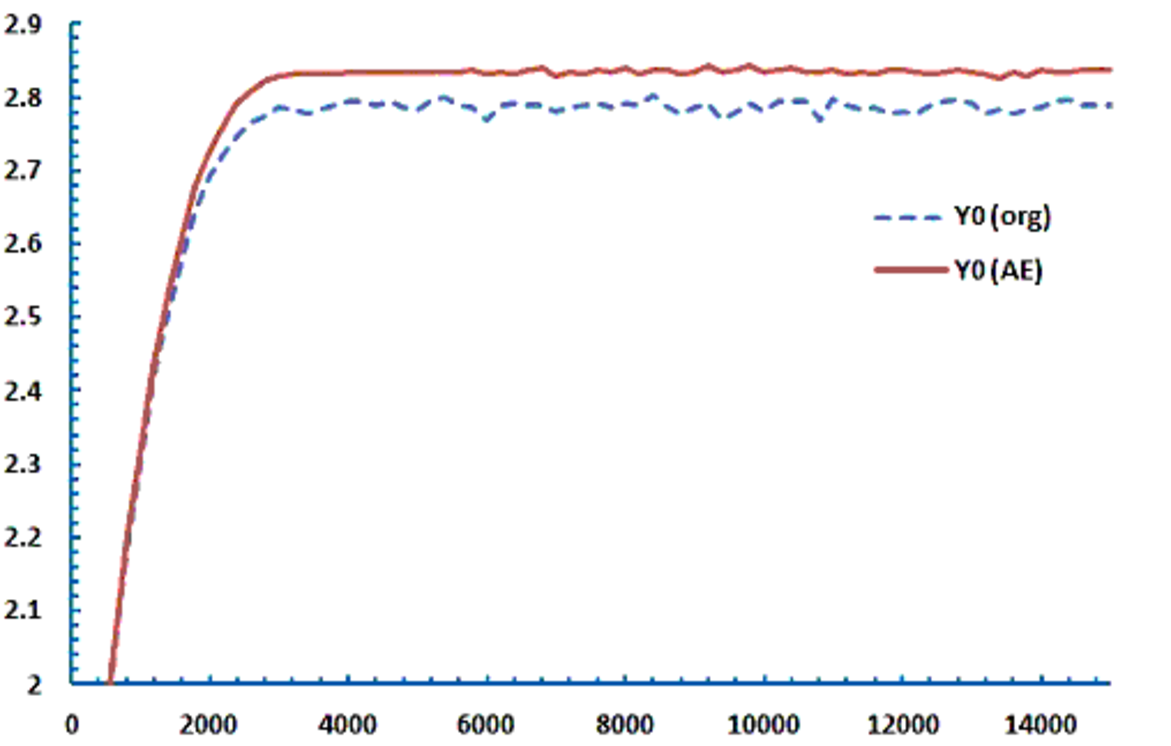}
\caption{\small Comparison of the loss function and the estimated $Y_0$ between
the direct use of the deep BSDE solver (indicated by a dashed line) and the one using AE
as prior knowledge (indicated by a solid line) for the 30-dimensional call spreads with correlation implied by $\gamma=0.06$
in (\ref{eq-rho}).}
\label{Fig-d30CSP}
\end{figure}

Finally, we provide the numerical results for a high dimensional setup with
\bea
\Phi(X_T)=\frac{1}{d}\sum_{i=1}^d \Bigl((X_T^i-K_1)^+-2(X_T^i-K_2)^+\Bigr), \nn
\eea
$\{d=30,(\mu^i)_{i=1}^d =0.05, r=0.01, R=0.06, (\sigma^i)_{i=1}^d=0.2, T=0.25, K_1=95, K_2=105\}$
and the same matrix $\rho$ with $\gamma=0.06$ given in the last subsection. 
The numerical comparison is given in Figure~\ref{Fig-d30CSP}.
Probably due to the diversification effects, one observes the quicker convergence of the estimated $Y_0$ for both cases.
However, the original deep BSDE solver without AE effectively ceases improvement  where 
the loss function is more than one magnitude larger than the case with AE.
It is likely that this is a cause of small gap in the estimated $Y_0$ given in the right-hand figure.

\section{An example of a reflected BSDE}
\subsection{Model}
We now study a reflected BSDE corresponding to the pricing problem of an American option.
The relevant BSDE where $Y_t$ corresponds to the option price is given by
\bea
&&Y_t=\Phi(X_T)-\int_t^T \Bigl\{ rY_s+\sum_{i,\alpha=1}^d Z_s^\alpha (\rho^{-1})_{\alpha,i}\frac{\mu^i+y^i-r}{\sigma^i}
\Bigr\}ds -\int_t^T \sum_{\alpha=1}^d Z_s^\alpha dW_s^\alpha+L_T-L_t, \nn \\
&&Y_t\geq \Phi(X_t),~t\geq 0, \quad \int_0^T [Y_t-\Phi(X_t)]dL_t=0,
\label{eq-RBSDE}
\eea
where $y^i>0$ is a dividend yield of the ith security $\{i=1,\cdots,d\}$, $(L_t,t\in[0,T])$ is the 
reflecting process that keeps the solution $Y_t$ from going below the barrier $\Phi(X_t)$ for every $t\in[0,T]$.
The other assumptions made in Section~\ref{sec-Bergman-model} are in force,
in particular, the dynamics of $(X_t)_{t\geq 0}$ is still given by (\ref{eq-X}).
The detailed derivation of the above BSDE is available, for example, in Chapter 6 of \cite{Zhang-text}.

Instead of using the penalization method~\cite{ElKaroui},
we extend the deep BSDE solver so that it learns the process $(L_t, t\in[0,T])$  directly in addition to $Y_0$ and $(Z_t,t\in[0,T])$.
We adopt the loss function
\bea
|\Phi(X_T)-\wh{Y}_T|^2+w\int_0^T \max(\Phi(X_t)-\wh{Y}_t,0)^2 dt \nn
\eea
where the weight $w:=2.0/T$ is used to take a balance between the terminal and the lateral conditions.
Remember that $\wh{Y}$ is the forwardly simulated process based on the estimated $Y_0$ and $(Z_t,L_t)_{t\in[0,T]}$.
We apply $dL_t$ to update the process $\wh{Y}$ only when $\wh{Y}_t\leq \Phi(X_t)$ so that we can avoid
the explicit inclusion of the second condition of (\ref{eq-RBSDE}) into the loss function.

\begin{remark}
Since it is impossible to make the loss function exactly zero, the weight factor $w$ slightly affects 
the estimated $Y_0$ (as well as the size of the loss function).
For the two examples we shall see below, halving the weight factor $w$ to $1.0/T$ lowers 
the American option prices by roughly 1.0\%.
\end{remark}

\subsection{Numerical examples}
\label{subsec-american}
In the following, we adopt
$\Phi(X_T)=\max\Bigl(\frac{1}{d}\sum_{i=1}^d X_T^i-K,0\Bigr)$
corresponding to an American basket call option.
The leading order asymptotic expansion is still given by
$Y_t^{(0)}=e^{-r(T-t)}\mbb{E}^{\mbb{Q}}\Bigl[\Phi(X_T)|\calf_t\Bigr]$ and $Z^{(0)}(=:Z^{\rm AE})$ as its deltas multiplied by $\sigma^i X^i$.
Although one cannot get the exact solution, it is not difficult to expand the solution in terms of $\sigma$
to obtain
\bea
Z^{\rm AE,\alpha}_t=N(d_c)\frac{1}{d}\sum_{i=1}^d e^{-y^i(T-t)}\sigma^i X_t^i\rho_{i,\alpha}+\calo(\sigma^2)~, \alpha=1,\cdots, d
\label{eq-ZAE-basket}
\eea
where
\bea
d_c&=&\frac{1}{\wt{\sigma}(t)\sqrt{T-t}}\Bigl(\frac{1}{d}\sum_{i=1}^d X_t^i e^{(r-y^i)(T-t)}-K\Bigr)\nn \\
\wt{\sigma}(t)&=&\frac{1}{d}\Bigl(\sum_{i,j=1}\sigma^i X_t^i e^{(r-y^i)(T-t)}(\rho\rho^\top)_{i,j}\sigma^j X_t^j 
e^{(r-y^j)(T-t)}\Bigr)^{1/2} ~.\nn
\eea

The above approximation  is based on the well-known {\it small-diffusion} expansion technique~\cite{Takahashi-review}.
We perturb the forward process $X$ as
\bea
&&dX_s^{i,\ep}=(r-y^i)X_s^{i,\ep}ds+\ep X_s^{i,\ep}\sigma^i \sum_{\alpha}\rho_{i,\alpha}dW_s^{\mbb{Q},\alpha}, s\geq t\nn \\
&&\quad X_t^{i,\ep}=X_t^i~. \nn
\eea
Notice the fact that the expansion of $X^{\ep}$ as a power series of $\ep$ is equivalent 
to that of $\sigma$ after setting $\ep=1$.

One sees that the 0th order expansion corresponds to the deterministic forward process 
$X_s^{i,(0)}=X_t^i e^{(r-y)(s-t)}, s\geq t$ and that the next order expansion is defined by
$dX_s^{i,1}=(r-y^i)X_s^{i,1}ds+X_s^{i,0}\sigma^i\sum_{\alpha}\rho_{i,\alpha}dW_s^{\mbb{Q},\alpha}, s\geq t$ and $X_t^{i,1}=0$.
Using the expansion up to the first order, it is easy to derive the result (\ref{eq-ZAE-basket}) since the process is now Gaussian.
Although one can continue the expansion to an arbitrary higher order,
it is expected to give only marginal effects when used in the deep BSDE solver.
As we can easily see from the above analysis, the leading order term of the small-diffusion expansion
always yield a Gaussian process regardless of the underlying process for $X$.
Therefore deriving $Z^{\rm AE}$ up to the first order of volatility can be performed easily in most cases.

\subsubsection{American call option}

\begin{figure}[h]
\hspace{-9mm}
\includegraphics[width=85mm]{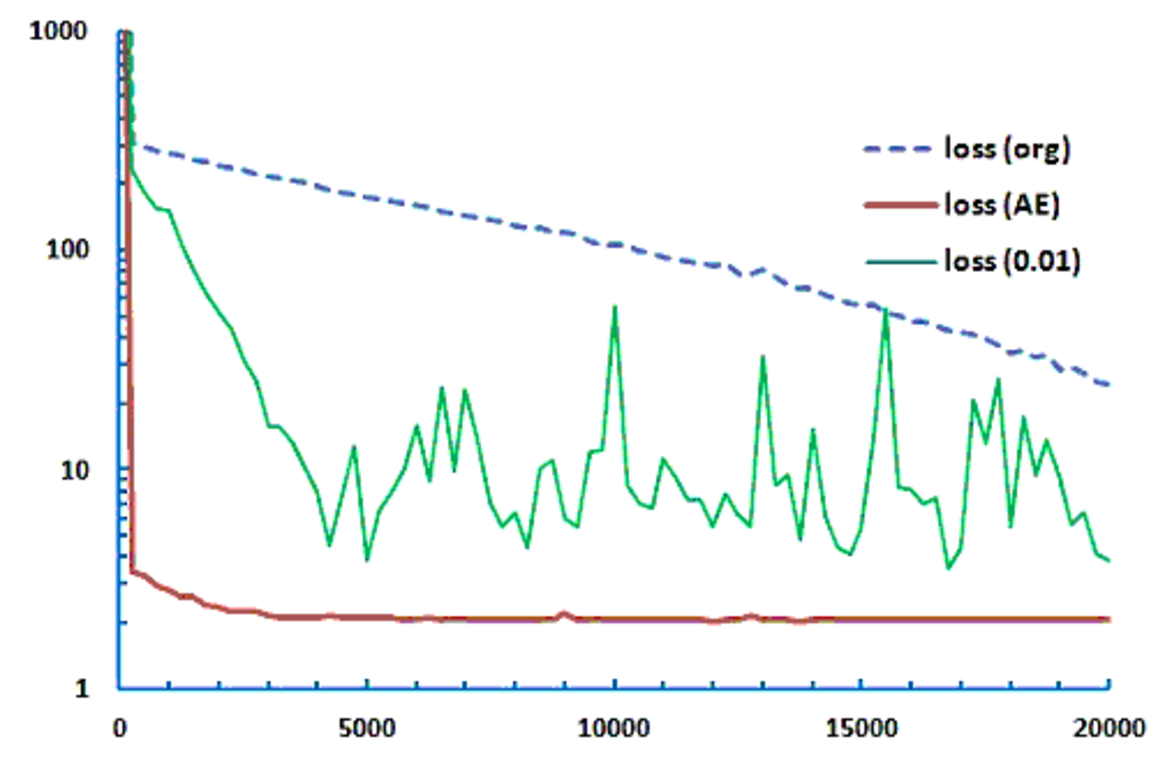}
\includegraphics[width=85mm]{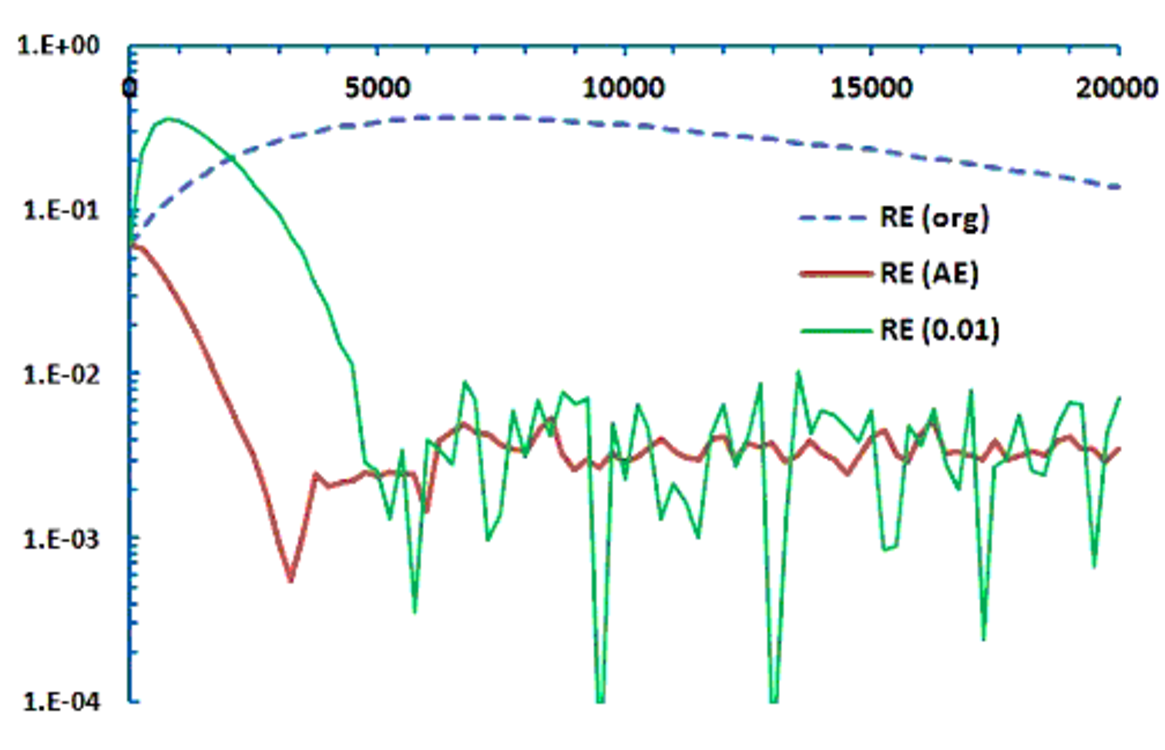}
\caption{\small Comparison of the loss function and the relative error between
the direct use of the deep BSDE solver and the one using AE
as prior knowledge for the 1-dimensional American call option.
For comparison, the case with learning rate $10^{-2}$ (without using AE ) 
is given by a green solid line.}
\label{Fig-d1-American}
\end{figure}

Let us first check the general performance by studying one-dimensional example.
We use $\{\mu=0.02,~y=0.07,~r=0.03,~T=0.5,~\sigma=0.2,~K=100,~x_0=110\}$. Note that the choice of $\mu$ should not affect $Y_0$.
The benchmark price based on a binomial tree model with 10,000 time steps available in the literature 
is $11.098$, while the corresponding European option price 
is $10.421$. The comparison of the loss function and the relative error is given in Figure~\ref{Fig-d1-American}.
In order to achieve an accurate estimate, fine discretization (n\_time$=100$) is used.
Since the direct use of the deep BSDE solver with the default learning rate yields very slow convergence,
we have also provided the case with the learning rate $10^{-2}$. 
The associated instability in the loss function as well as $Y_0$ suggests that one needs to fine-tune
the learning rate dynamically in the deep BSDE solver for achieving the comparable performance
to the case with AE. 

\subsubsection{American 50-dimensional basket call option}
We now study a 50-dimensional American basket call option.
Let us use $\{\mu^i=0.02, y^i=0.07, x_0^i=110, \sigma^i=0.2\}_{i=1}^{50}$,
$K=100,~T=0.5,~r=0.03$ and $\gamma=0.07$ in (\ref{eq-rho}), which implies around 30\% correlation 
for every pair of $X$'s. We have used  (n\_time$=100$) time partition as before.
The price of the corresponding European option is estimated as $8.46$ by a simulation with 500,000 paths.

\begin{figure}[h]
\hspace{-9mm}
\includegraphics[width=85mm]{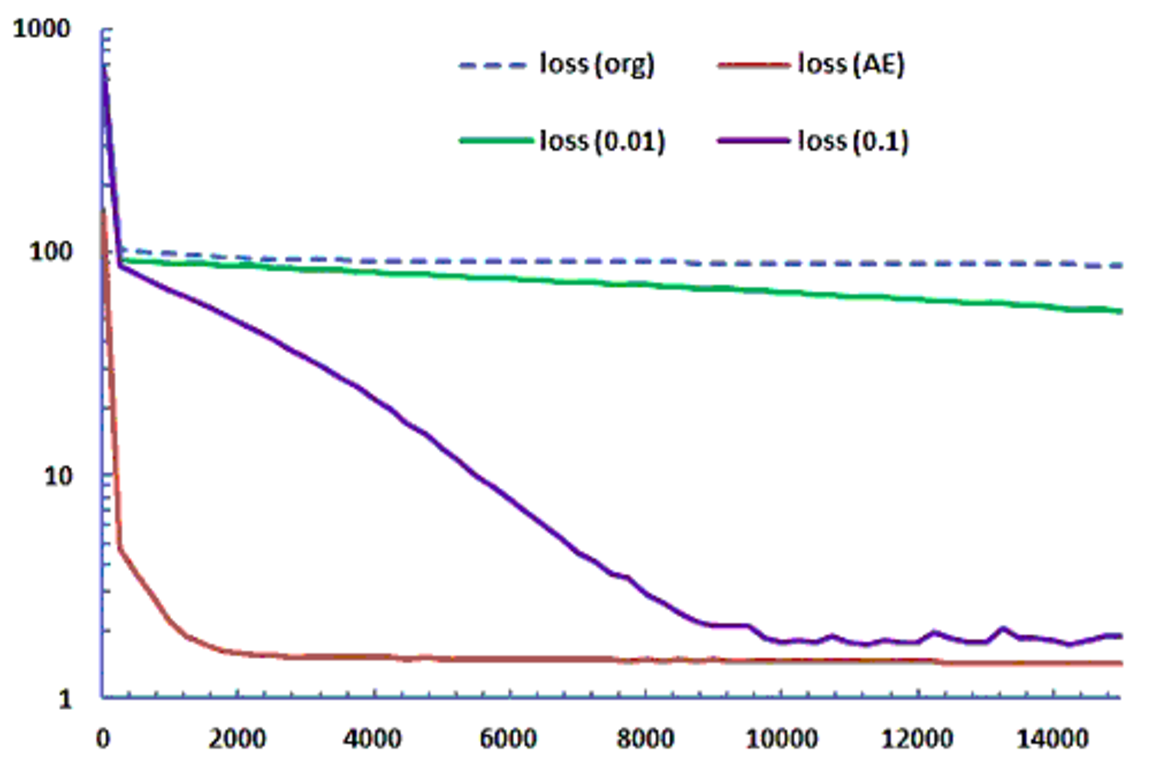}
\includegraphics[width=85mm]{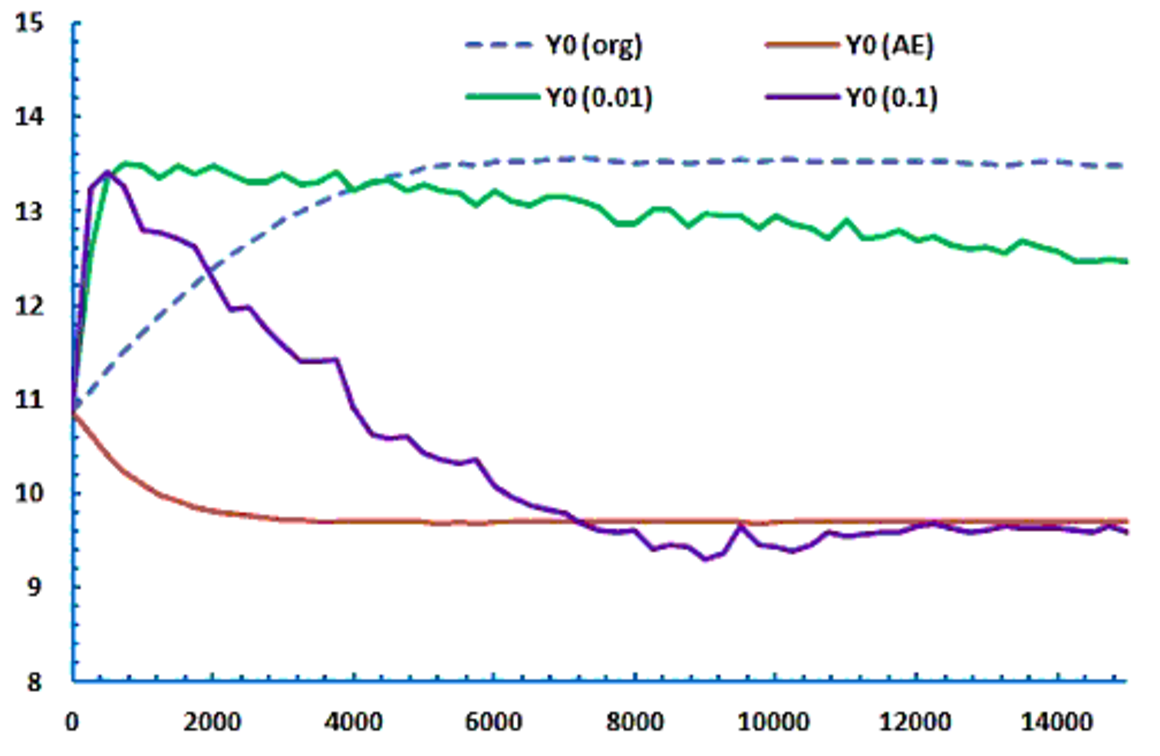}
\caption{\small Comparison of the loss function and the estimated option price $Y_0$ between
the direct use of the deep BSDE solver and the one using AE
as prior knowledge for the 50-dimensional American basket call option.
For comparison, the cases with learning rate $10^{-2}$ and $10^{-1}$ (without using AE ) 
are also given by green and purple lines.}
\label{Fig-d50-American}
\end{figure}

As one can see from Figure~\ref{Fig-d50-American}, the convergence is very quick when AE is used as prior knowledge.
After 2,000 iterations, $Y_0$ is settled around 9.7. When AE is not used, the convergence is very slow.
Even if we use an extremely large learning\_rate$=0.1$, it takes more than 10,000 iterations to give a comparable size of loss function.
The stability of estimated $Y_0$ with AE clearly stands out from the others.
In the deep BSDE solver, we estimate $(Z_t, L_t)$ {\it at each time step} using neural networks with multiple layers.
Therefore quick convergence brought by a simple AE approximation is a great advantage, in particular, for the problems that require many time-steps for accurate estimates. Comparing to the previous one-dimensional example, one clearly sees that the optimal choice 
of the learning rate depends on the details of the settings (such as, the number of assets and the correlation among them) 
even for the BSDE with the same form.

\begin{remark}
Changing the code for the penalization method~\cite{ElKaroui} is very simple.
The solution of the penalized BSDE, which is obtained by replacing $L_T-L_t$ with
\be
\frac{1}{\ep}\int_t^T \max(\Phi(X_s)-Y_s,0)ds~, \nn
\ee
is known to converge to that of (\ref{eq-RBSDE}) in the limit of $\ep\downarrow 0$. Although there is no need to estimate the process $L$,
we have found that the numerical results depend quite sensitively on the size of $\ep$.
Moreover small $\ep$ slows down the learning process significantly and hence not useful in many cases.
It helps, however, double checking the correct implementation by comparing the numerical results.
\end{remark}

\section{An example of a Quadratic BSDE}
\subsection{Model}
We now consider the following qg-BSDE
\bea
Y_t=\Phi(X_T)+\int_t^T \frac{a}{2}|Z_s|^2-\int_t^T \sum_{\alpha=1}^d Z_s^\alpha dW_s^\alpha,~t\in[0,T]
\label{eq-qgBSDE}
\eea
where $a\in \mbb{R}$ is a constant, $W$ a $d$-dimensional Brownian motion, 
$\Phi:\mbb{R}^d\rightarrow \mbb{R}$ is a bounded Lipschitz continuous function.
For simplicity, we assume that the associated forward process is given by
\bea
X_t^i=x_0+\int_0^t \sigma X_s^i \sum_{\alpha=1}^d \rho_{i,\alpha}dW_s^\alpha, ~t\in[0,T], ~i=1,\cdots,d  \nn
\eea
with a common initial value $x_0>0$, and a volatility $\sigma>0$. $\rho=(\rho_{i,j})_{i,j=1}^d$
is a square root of correlation matrix among $X^i$ and  assumed to be invertible.

Thanks to this special form, it is easy to derive a closed form solution
\bea
Y_t=\frac{1}{a}\log\Bigl(\mbb{E}\Bigl[\exp\bigl(a \Phi(X_T)\bigr)\bigr|\calf_t\Bigr]\Bigr),~t\in[0,T]
\label{eq-qgBSDE-solution}
\eea
by applying It\^o formula to $e^{a Y_t}$. The existence of a closed-form solution
allows us to test the performance of the deep BSDE solver.
Note however that the numerical evaluation of qg-BSDE is known to be very difficult.
See discussions in Imkeller \& Reis (2010)~\cite{Imkeller},
Chassagneux \& Richou (2015)~\cite{Richou} and Fujii \& Takahashi (2018)~\cite{FT-short-term}.

A formal application of the method \cite{FT-analytic} to the current case gives
$Y_t^{(0)}=\mbb{E}\Bigl[\Phi(X_T)|\calf_t\Bigr]$ as the leading order asymptotic expansion,
and hence $Z_t^{(0)} (=:Z_t^{\rm AE})$ can be derived as deltas in exactly the same manner as in the last section.
Although the asymptotic expansion methods in \cite{FT-analytic, Takahashi-Yamada}
are only proved for the Lipschitz BSDEs, using Malliavin's differentiability 
and the associated representation theorem for $Z$ given in Ankirchner, Imkeller \& Dos Reis (2007)~\cite{Imkeller-Reis},
one can justify the method also for the quadratic case in a similar way.
The asymptotic expansion for the Lipschitz BSDEs with jumps in \cite{FT-AE} can also be extendable to a quadratic-exponential
growth BSDEs by using the results of Fujii \& Takahashi (2018)~\cite{FT-Qexp}.
The details may be given in different opportunities.

\subsection{Numerical examples}

\begin{figure}[h]
\hspace{-9mm}
\includegraphics[width=85mm]{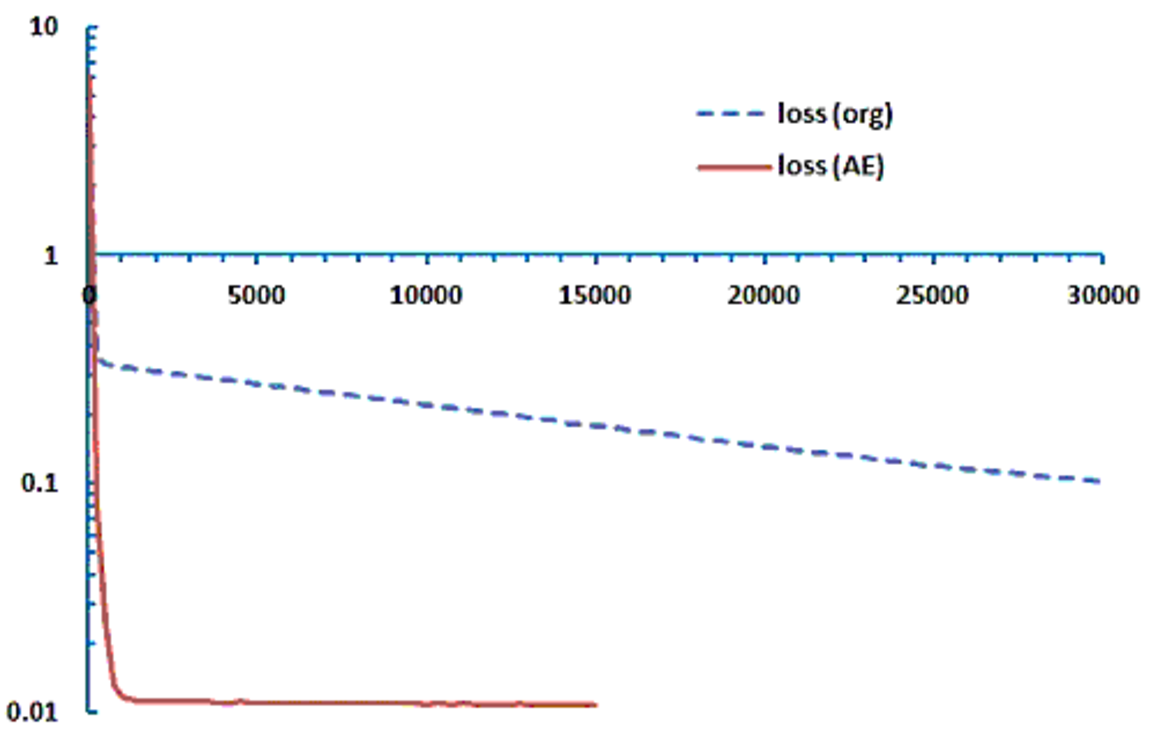}
\includegraphics[width=85mm]{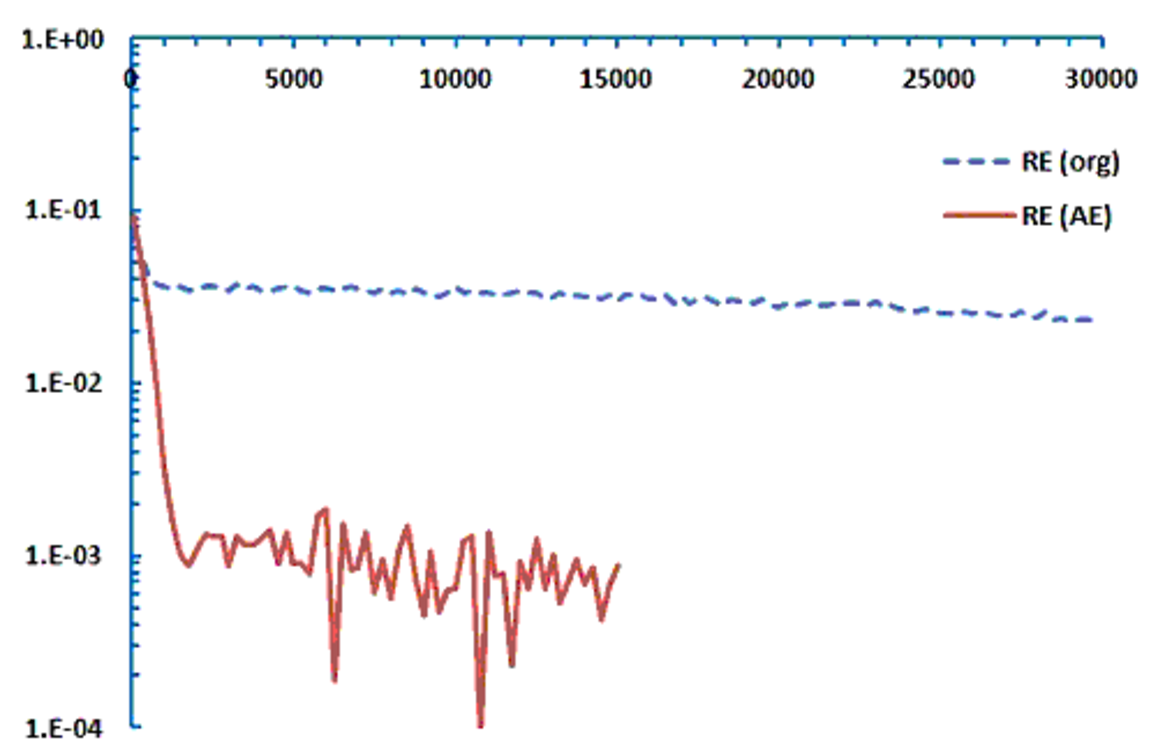}
\caption{\small Comparison of the loss function and the relative error between
the direct use of the deep BSDE solver (indicated by a dashed line) and the one using AE
as prior knowledge (indicated by a solid line) for the 50-dimensional model with zero correlation.}
\label{Fig-d50-qg0}
\end{figure}

We suppose a bounded terminal condition defined by
\bea
\Phi(X_T)=\frac{1}{d}\sum_{i=1}^d \Bigl(\max(X_T^i-K_1,0)-\max(X_T^i-K_2,0)\Bigr)
\eea
with two constants $0<K_1<K_2$.
As a first example, we have tested a 50-dimensional model with zero correlation: 
${\rm set_0}:=\{d=50, a=1.0,  T=0.25, K_1=95, K_2=105, \sigma=0.2, x_0=100, \rho=\bold{I}_{d\times d}\}$.
The solution  (\ref{eq-qgBSDE-solution}) is estimated as $Y_0=5.01$ by a simulation with one million paths.
We use $\text{n\_time}=25$ as discretization. In Figure~\ref{Fig-d50-qg0}, we have compared the performance of the deep BSDE solver
with and without AE as prior knowledge.
When the asymptotic expansion is used, the convergence is achieved just after a few thousands iterations and 
the relative error becomes $\leq$0.1\%. On the other hand, the learning process proceeds very slowly 
when the deep BSDE solver is directly applied without using AE. Even after 30,000 iterations, both of the loss function 
and the relative error are still larger than the former by more than an order of magnitude. 
It seems that clever dynamic tuning of the learning rate is necessary for achieving comparable speed of convergence and stability 
to those for the case with asymptotic expansion.

\begin{figure}[h]
\hspace{-9mm}
\includegraphics[width=85mm]{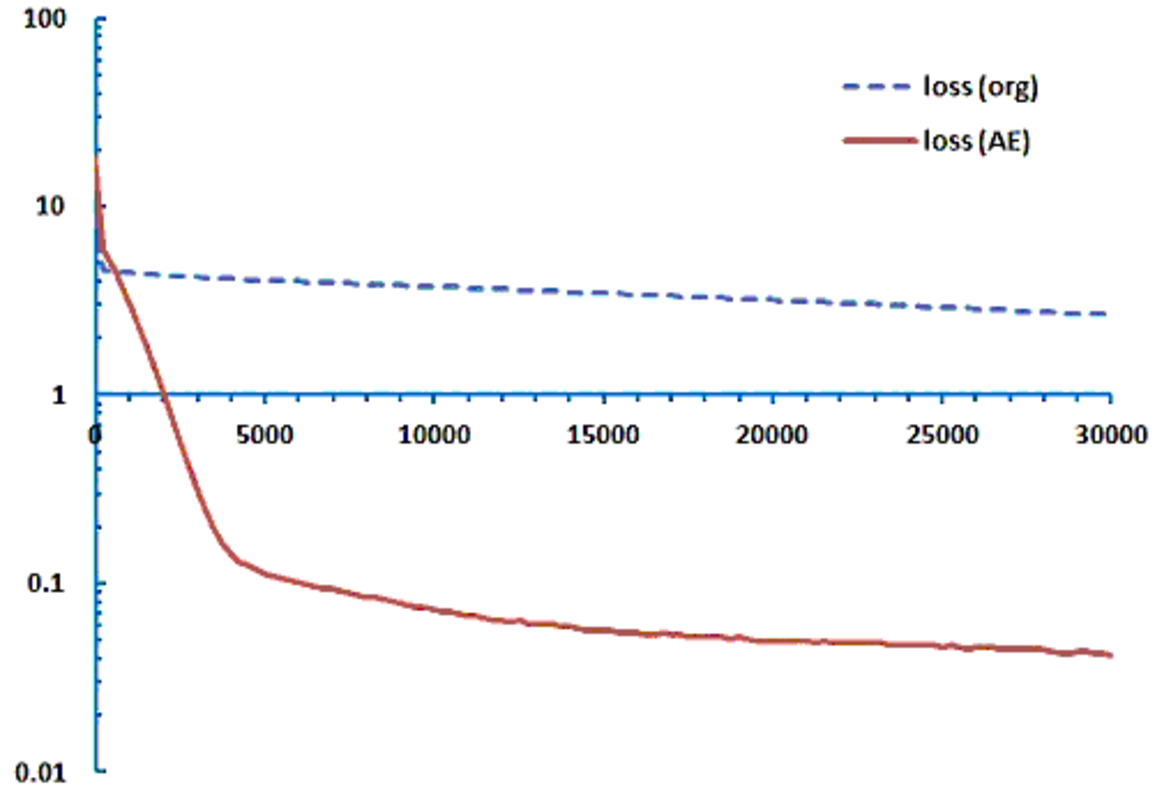}
\includegraphics[width=85mm]{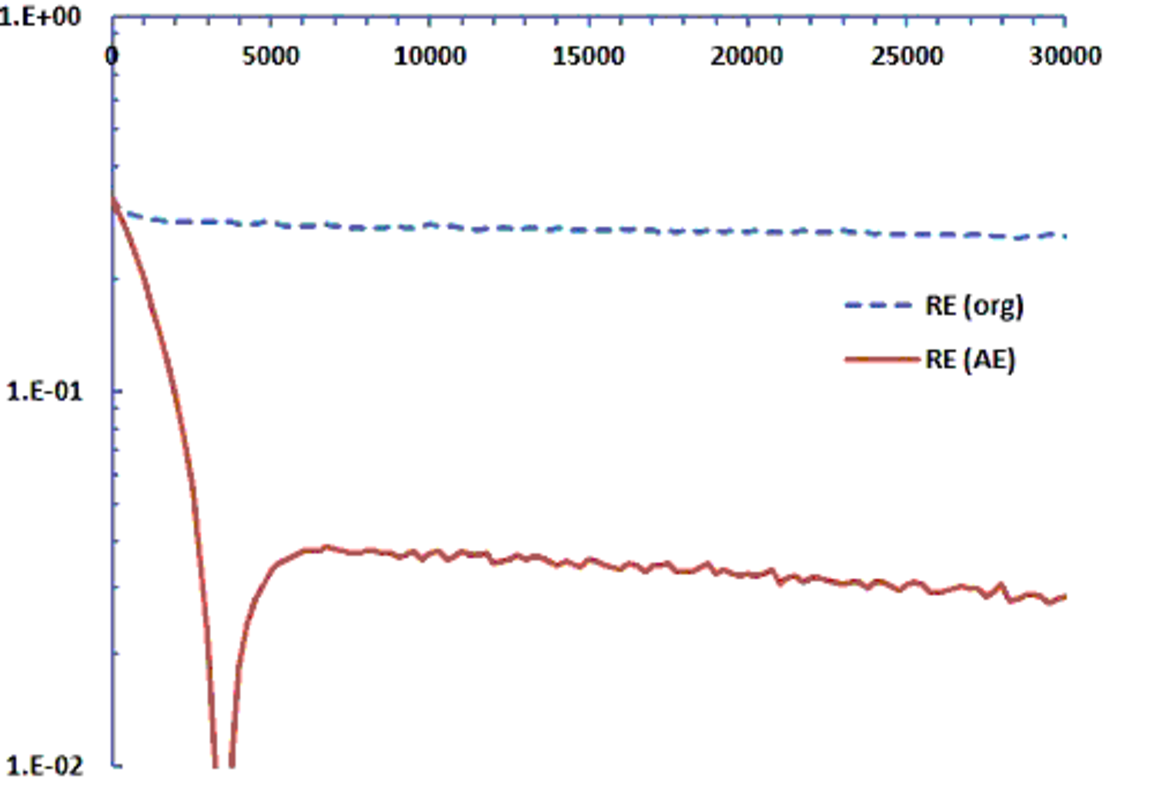}
\caption{\small 
Comparison of the loss function and the relative error between
the direct use of the deep BSDE solver (indicated by a dashed line) and the one using AE
as prior knowledge (indicated by a solid line) for the 50-dimensional model with correlation implied by $\gamma=0.07$
in (\ref{eq-rho}).}
\label{Fig-d50-qgCorr}
\end{figure}

Next, we have studied the impact of correlation among $X$'s.
Since the regressors have non-zero correlation, one can expect that the learning process becomes more time consuming.
We have used the same parameters in $\rm set_0$ except that the matrix $\rho$ is now defined by (\ref{eq-rho}) with $\gamma=0.07$,
which implies about 30\% correlation among $X$'s.
In this case, the solution (\ref{eq-qgBSDE-solution}) estimated by one million paths is $Y_0=6.78$.
The comparison of the performance is given in Figure~\ref{Fig-d50-qgCorr}.
Although the accuracy is deteriorated in the both cases, the deep BSDE solver with the asymptotic expansion still achieves
the relative error $3\sim 4\%$ after 5,000 iterations. When AE is not used, the relative error remains more than $25\%$
even after 30,000 iteration steps. Similarly to the previous examples, one sees that 
the use of AE as prior knowledge effectively
ameliorates the problem of correlated inputs also for this case.

\begin{figure}[h]
\hspace{-9mm}
\includegraphics[width=85mm]{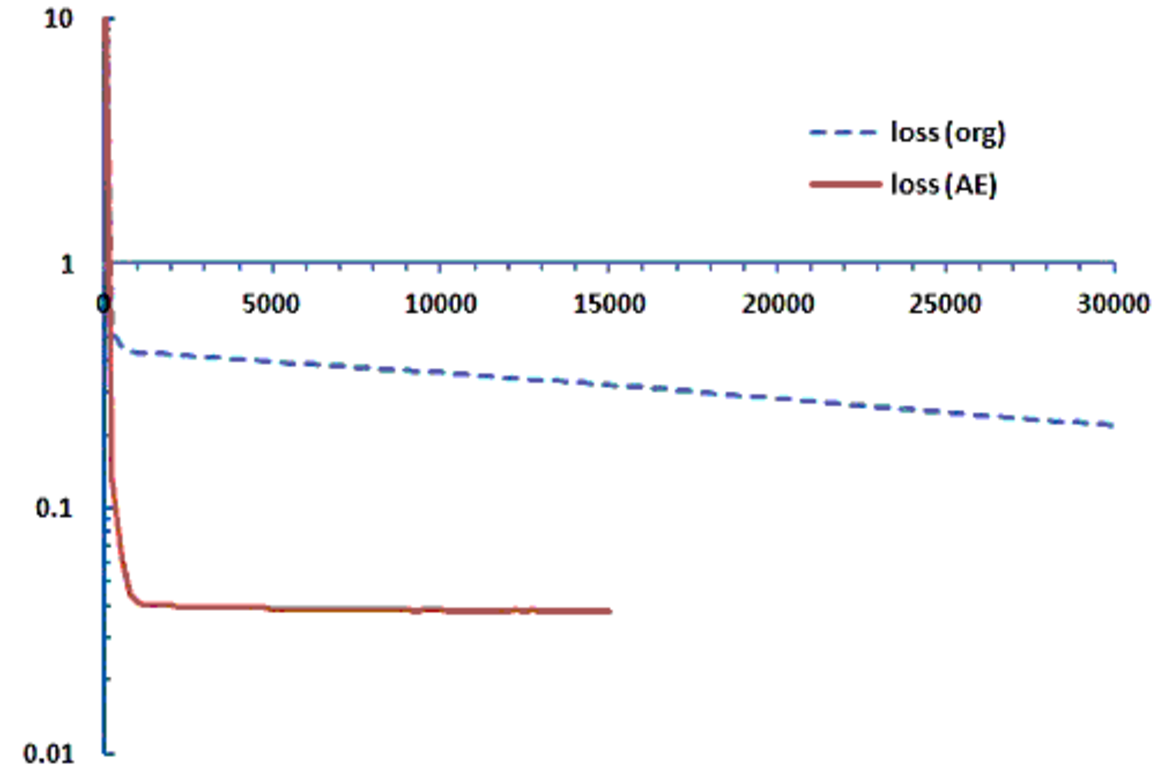}
\includegraphics[width=85mm]{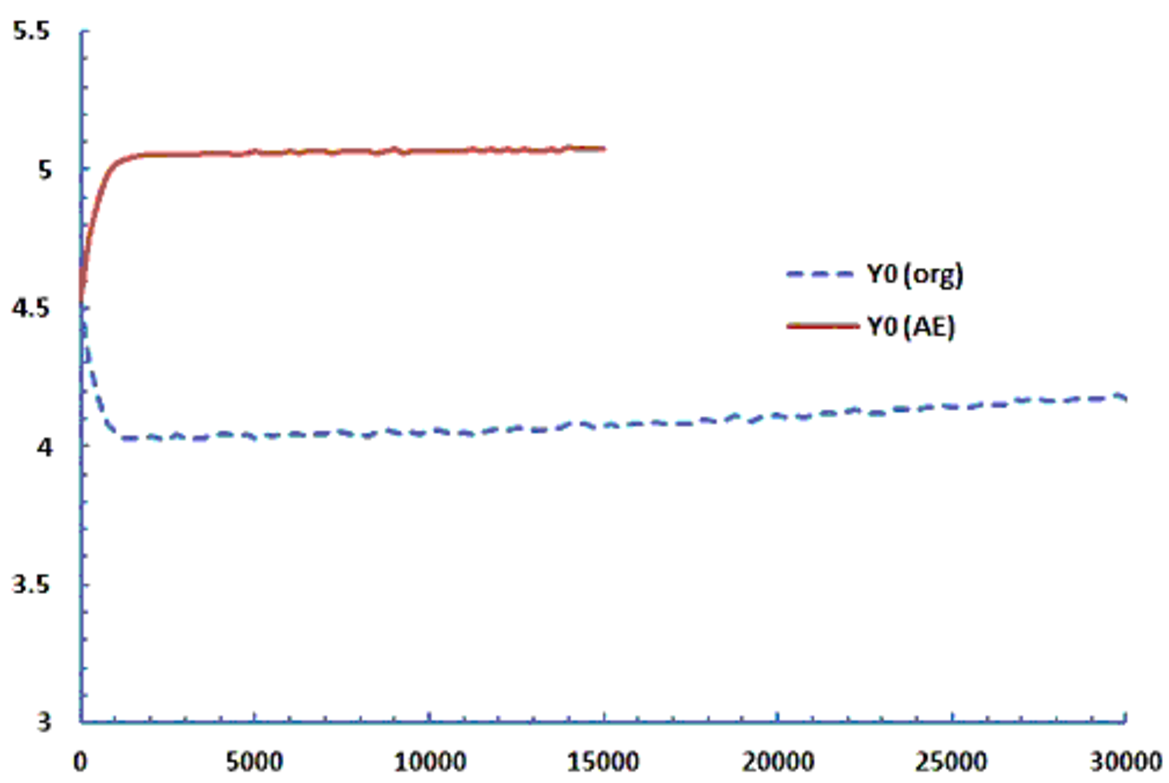}
\caption{\small 
Comparison of the loss function and the estimated $Y_0$ between
the direct use of the deep BSDE solver (indicated by a dashed line) and the one using AE
as prior knowledge (indicated by a solid line) for the 50-dimensional model with $a=5.0,~\sigma=1.0$.}
\label{Fig-d50-qgBa}
\end{figure}

Finally, let us study a bit extreme situation with a large quadratic coefficient as well as volatility.
We set $\{d=50, a=5.0,  T=0.25, K_1=95, K_2=105, \sigma=1.0, x_0=100, \rho=\bold{I}_{d\times d}\}$
and increase the number of time partition to n\_time$=50$.
The solution (\ref{eq-qgBSDE-solution}) estimated by a million paths of Monte Carlo simulation with the same step size
is given by $Y_0=5.17\pm 0.01$. We have provided the numerical results in Figure~\ref{Fig-d50-qgBa}.
Although the loss function becomes larger by a factor of few, the deep BSDE solver with AE
quickly reaches its equilibrium after a few thousands steps and the relative error is around $2\%$.
As is clearly seen from the graph, the estimated $Y_0$ obtained without using AE 
is quite far from the target after 30,000 iterations steps.

\section{Concluding remarks}
In this work, we have demonstrated that one can greatly accelerate the learning process by using a 
simple approximation formula as ``prior knowledge". 
This overcomes the issue of slow progress of 
the learning process in the presence of the non-smooth functions as well as correlated security processes
for the deep BSDE solver, and may pave the way for the practical use of BSDEs 
with more realistic description of non-linearity in the financial markets.
As a nature of machine-learning technique, accelerated convergence 
is expected to be a generic phenomenon regardless of the exact form of algorithm
when appropriate prior knowledge is given. 

Although appropriately chosen hyper-parameters may achieve quicker convergence, 
their optimization is usually a difficult task requiring trial and error.
For example,  TensorFlow provides a simple tool to make the learning rate decay at a certain rate,
but choosing an appropriate {\it decay rate}  becomes another trouble because we do not know, {\it a priori},  
the ``limit" of the loss function, which is determined by the size of the delta-hedging error unavoidable for given discretization.
If the speed of decay is too fast, then there remains 
large loss function at the time when the learning rate becomes very small. This results in slow convergence.
On the other hand, if the speed of decay is too slow, one has to wait unnecessary long time 
for the learning rate becoming small enough to yield a stable estimate. A simple AE formula nicely 
solves these issues. Moreover, it may be combined with the hyper-parameter optimization 
to enhance its performance.

Application of the deep learning methods to the BSDEs with jumps remains as an important
challenge. Relatively small intensity of the jumps 
(such as those in credit models) is expected to make the learning process very hard to proceed.
An analytic approximation of the  jump coefficients available by the asymptotic expansion in \cite{FT-AE}
may mitigate the difficulty.

\section*{Acknowledgement}
The research is partially supported by Center for Advanced Research in Finance (CARF).



\end{document}